\documentclass{emulateapj}
\usepackage{graphicx}
\usepackage{amssymb}
\usepackage{amsmath}
\usepackage{ulem}
\usepackage[colorlinks=true,linkcolor=blue,anchorcolor=blue,citecolor=blue,urlcolor=blue]{hyperref}  
\usepackage{array}
\newcolumntype{P}[1]{>{\centering\arraybackslash}p{#1}}
\usepackage{comment}

\usepackage[flushleft]{threeparttable}

\shorttitle{Galaxies as High-Resolution Telescopes}
\shortauthors{Barnacka}

\begin{document}
\title{Galaxies as High-Resolution Telescopes}

\author{Anna Barnacka$^{1,2}$}

\affil{$^1$Harvard-Smithsonian Center for Astrophysics, 60 Garden St, MS-20, Cambridge, MA 02138, USA\\
$^2$Astronomical Observatory, Jagiellonian University, Cracow, Poland \\ }

\email{abarnacka@cfa.harvard.edu}

\begin{abstract}
Recent observations show a population of active galaxies with milliarcseconds offsets between optical and radio emission. 
Such offsets can be an indication of extreme phenomena associated with supermassive black holes including 
relativistic jets, binary supermassive black holes, or even recoiling supermassive black holes. 
However, the multi-wavelength structure of active galaxies at a few milliarcseconds  cannot be resolved with direct observations. 
We propose using strong gravitational lensing to elucidate the multi-wavelength structure of sources. 
When sources are located close to the caustic of lensing galaxy, 
 even small offset in the position of the sources
results in a drastic difference in the position and magnification of mirage images. 
We show that the angular offset in the position of the sources can be amplified more than 50 times in the observed position of mirage images.
We find that at least 8\% of the observed gravitationally lensed quasars will be in the caustic configuration. 
The synergy between SKA and Euclid will provide an ideal set of observations for thousands of gravitationally lensed sources in the caustic configuration, 
which will allow us to resolve the multi-wavelength structure for a large ensemble of sources, and study the physical origin of radio emissions, their connection to supermassive black holes, and their cosmic evolution.

\end{abstract}


\section{Introduction}

The inner regions of active galaxies can host the most extreme and energetic phenomena in the universe including
relativistic jets, binary supermassive black holes, or recoiling supermassive black holes
 \citep{1980Natur.287..307B,1984RvMP...56..255B,2002ApJ...565..244H,2003ApJ...582..559V,2006AIPC..856....1M,2009ApJ...698..956C,2011MNRAS.412.2154B,2012Sci...338..355D,2017arXiv170306143B,2016ApJ...830...50M,2016arXiv161100554D,2016A&A...588A.125R,2017MNRAS.464.3131K,2017arXiv170604010P}. 
At scales smaller than a few hundred parsecs 
the VLBI technique is needed to provide angular resolution sufficient to resolve the sources.
Thus, it is commonly believed that the majority of radiation from these powerful sources originates within 
a milliarcsecond ($\sim10\,$parsecs) of the supermassive black hole \citep{2011MNRAS.417L..11S,2014ApJ...789..161N,2013ApJ...768...54B,2014A&A...567A.113B}. 

However, observations provide an evidence that the radio core that is assumed to be among the closes structures to the supermassive black hole  can be located $\sim10\,$pc 
or more from it \citep{2011arXiv1110.6463A,2012IJMPS...8..356J,2012arXiv1204.6707M,2012arXiv1201.5402M}. 
Recent comparisons of VLBI positions of extragalactic radio-loud sources 
with their optical counterparts from  the first Gaia release \citep{2016A&A...595A...4L} 
shows that the radio and optical emission regions are physically offset for a fraction of luminous AGN 
\citep{2016A&A...595A...5M,2017A&A...598L...1K,2017ApJ...835L..30M,2017arXiv170407365P,2017MNRAS.467L..71P}. 
The presence of $\sim10\,mas$ offsets in the radio-optical positions of reference AGN
was previously investigated by \citet{2013A&A...553A..13O,2014AJ....147...95Z}.

Moreover, \citet{2016ApJ...821...58B} demonstrated that the observed radio core can be located more 
than $6\,mas$ from the supermassive black hole. 
Such large offsets cannot be explained by a frequency dependent shift of the position of the radio core, known as the core shift effect \citep{1979ApJ...232...34B,1981ApJ...243..700K,2008A&A...483..759K}.
The core shift effect predicts offsets of less than $1\,mas$ \citep{2011A&A...532A..38S}. 
It is unknown how often the observed radio core is offset from the central engine, 
what is the physical origin of the offset,
or what the offset distribution is. 

Here, we propose using strongly lensed systems located close to the caustic of the lensing galaxy to investigate the multi-wavelength structure of sources at a few $mas$ scale. 
The caustic of the lensing galaxy is a place where mirage images merge or are created. 
The proximity of the caustic results in drastic changes of position and magnification of the mirage images for even small offsets in the source position. 
Thus, if optical and radio emissions originate from the same region, 
the position of the mirage images observed using optical and radio telescopes will be coincident. 
However, if there is even a small offset between optical and radio emissions, then the positions of mirage images will differ significantly. 
We provide a brief definition of the caustic in Appendix~\ref{sec:Caustic}.

In Section~\ref{sec:ToyModel}, we investigate a toy model with an example of offset sources located close to the caustic. 
In Section~\ref{sec:MC}, we use Monte Carlo Simulations to explore the offset amplification for the entire region along the caustic. 
In Section~\ref{sec:CausticProbability}, we discuss the probability that the lensed source will be located close to the caustic of the lensing galaxy. 
In Section~\ref{sec:Astrometry}, we point out a solution for the astrometry.
In Section~\ref{sec:Discussion}, we discuss 
the implication of the offsets for the origin of the emission, 
we discuss lens modeling, 
and synergy between SKA and Euclid surveys.
We conclude in Section~\ref{sec:Conclusions}.
We present a brief introduction to the gravitational lensing formalism
and the properties of  a caustic of an elliptical lens in Appendices~\ref{sec:Formalism} and~\ref{sec:Caustic}.

\begin{figure}
\begin{center}
\includegraphics[width=10.cm,angle=0]{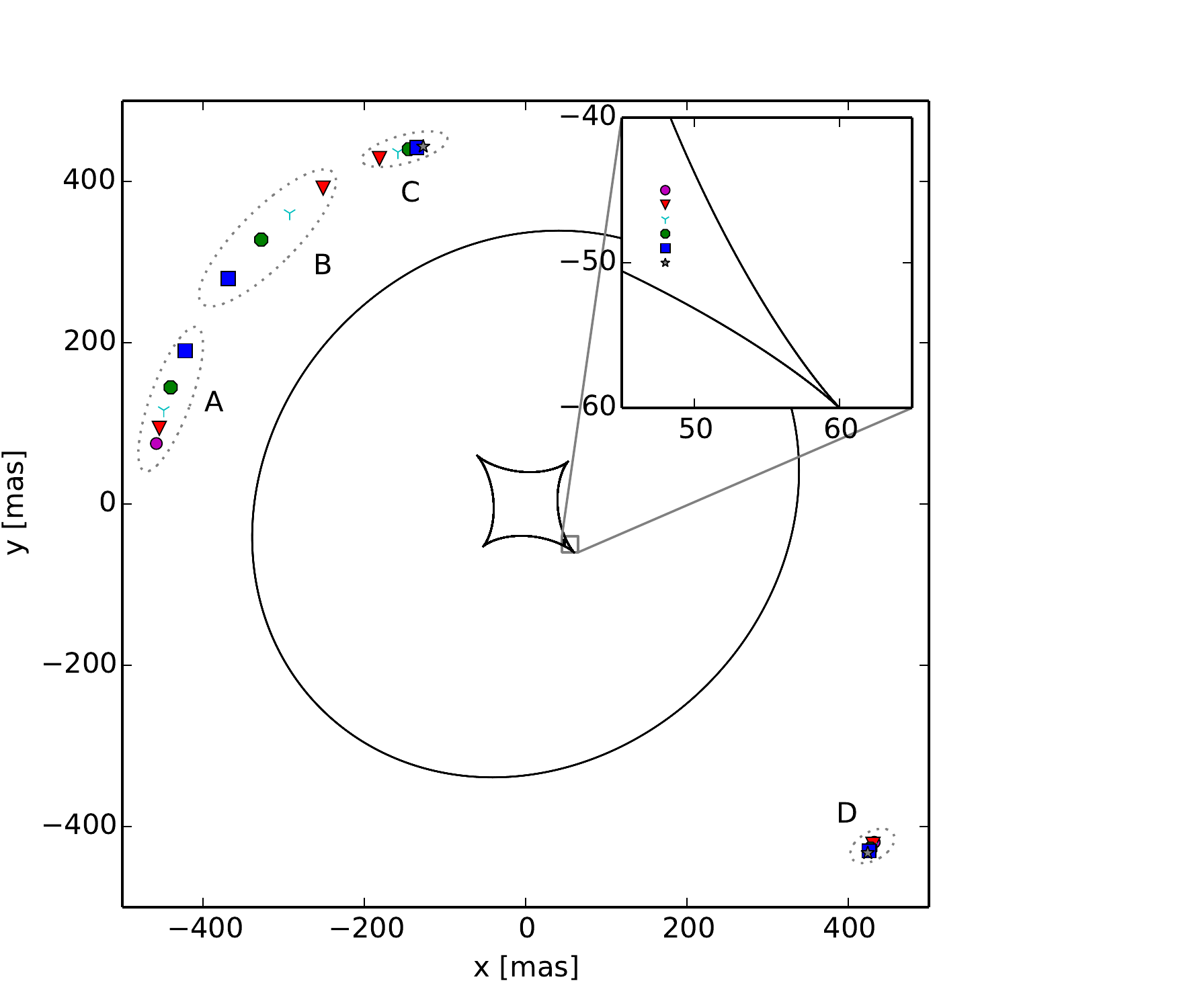}
\end{center}
\caption{\label{fig:ImageSeq} 
		       An illustration of mirage images formation by an elliptical lens.  
		       Black lines show the caustic curves of the lensing galaxy, which form an outer ellipse and an inner diamond shapes in the source plane. 
		       Six emitting regions are located close to the inner caustic. 
		       The 18 times zoomed source region is shown in the top-right corner.
		       The emitting regions in the source plane are separated by 1  to  $5\,mas$.
                          $1\,mas$ corresponds to $\sim10\,$pc in the source plane. 
                          The lens produces multiple images of each emitting region. 
                          The emitting regions in the source plane are presented using the same colors and point styles as their mirage images in the lens plane. 
                          Grey dotted ellipses group classes of the mirage images to facilitate a comparison of the image positions.
                          The most drastic changes are visible for the B image class where the positions differ by $\sim50\,$mas 
                          for emitting regions offset by  $1\,mas$ in the source plane.
                          }
\end{figure}

\section{Toy Model: Source Close to the Caustic }
\label{sec:ToyModel}

We start with a toy model consisting of six aligned emitting regions separated by $1\,mas$. 
Such configuration represents different offsets from 1 to  $5\,mas$ for sources offset by  $1\,mas$,
or can imitate a jet consisting of six distinct emission components. We refer to these emitting regions as knots.
We choose $1\,$mas distance to represent the offset expected between the low-frequency radio (2 GHz) 
and the optical emission region caused by the core shift effect \citep{2011A&A...532A..38S}.  
The larger distance of $5\,$mas was chosen to examine offsets found using the first Gaia release. 

We place the toy model sources close to the caustic (see the top-right corner in Figure~\ref{fig:ImageSeq}).
In our simulations, the source is at redshift 2. 
We put a lensing galaxy at redshift 0.5. 
We model the lens as an Singular Isothermal Ellipsoid (SIE; see Appendix~\ref{sec:Caustic}) with an ellipticity of $e=0.2$, 
an angle $\phi=45^\circ$, and the velocity dispersion $\sigma =170\,km/s$.
In the considered configuration, the Einstein ring radius is 0.5". 

For each knot, we find positions of the mirage images and their magnifications. 
Figure~\ref{fig:ImageSeq} shows the sequence of mirage images for six knots.
Remarkably, the positions of the mirage images differ by $20-50\,mas$  even for sources separated by only $1\,mas$. 

Figure~\ref{fig:ImageSeq}  shows that the largest variation in the positions of the images occurs for the B images. 
In the considered configuration, 
the angular offset in the source plane as small as $1\,mas$ can result in an offset between positions of the mirage images as large as $50\,$mas. 

We calculate the total change in the position of mirage images $\Delta \theta$ between two sources
by summing offsets between all mirage images. 
In our toy model, the mirage images of sources separated by $\Delta \beta =1\,mas$ 
are offset by $\Delta \theta \sim100\,mas$. 
Thus, the offset amplification\footnote[1]{Offset Amplification is defined as a ratio between an angular offset between mirage images, $\Delta\theta$,  
to the offset in the source plane $\Delta\beta$}  is $\sim100$. 
In addition to the change in the position of mirage images, there is a significant change in flux magnification.
The magnification ratio between mirage images can be used as an additional way to elucidate the spatial origin of the source. 

If we consider Knots~2 and~5 (red triangles and blue squares in Figure~\ref{fig:ImageSeq}),  
the offset  between these knots in the source plane is $\Delta \beta = 3\,mas$.
Remarkably, the positions of the mirage images  differ by $\Delta \theta \sim300\,mas$.
Thus, the offset amplification can allow us to determine the offset between radio and optical emission using existing telescopes.

\section{Monte Carlo Simulations }
\label{sec:MC}

To evaluate the offset amplification\footnotemark and flux magnification 
for pairs of offset sources at different orientations  and locations with respect to the caustic we run Monte Carlo simulations. 
We use the lens configuration as described in Section~\ref{sec:ToyModel}. 

The total magnification of the source as a function of the source position is shown in Figure~\ref{fig:TotalMag}. 
The total magnification is a sum of magnifications of all mirage images for given source positions. 
Inside the inner caustic, four magnified images are created and the total magnification is greater than 10.
When a source approaches a side of the caustic, the total magnification is greater than 100. 
At the caustic, two of the mirage images merge. 
Thus, the number of images changes.
The total magnification of a source outside the caustic is of the order of a few. 

The image magnification is determined by the second derivative of the effective potential  (see Appendix~\ref{sec:Formalism}). 
The positions of mirage images, and thus the offsets, are determined by the first derivative of the effective potential.
As a result, image magnification is changing faster that image position when a source is located close to the caustic.  
However, the advantage of using the image positions to reconstruct the origin of 
the emission is that the image positions are less sensitive to the substructures in the mass distribution of the lens and the effects of external shear.  

We randomly select $10^6$ positions in the source plane within 0.25 Einstein radius from the center of the lens. 
Such a radius allows us to include the entire inner caustic area. 
The position and magnification of mirage images are obtained using {\tt glafic} code \citep{2010PASJ...62.1017O}.

\begin{figure}
\begin{center}
\includegraphics[width=9.cm,angle=0]{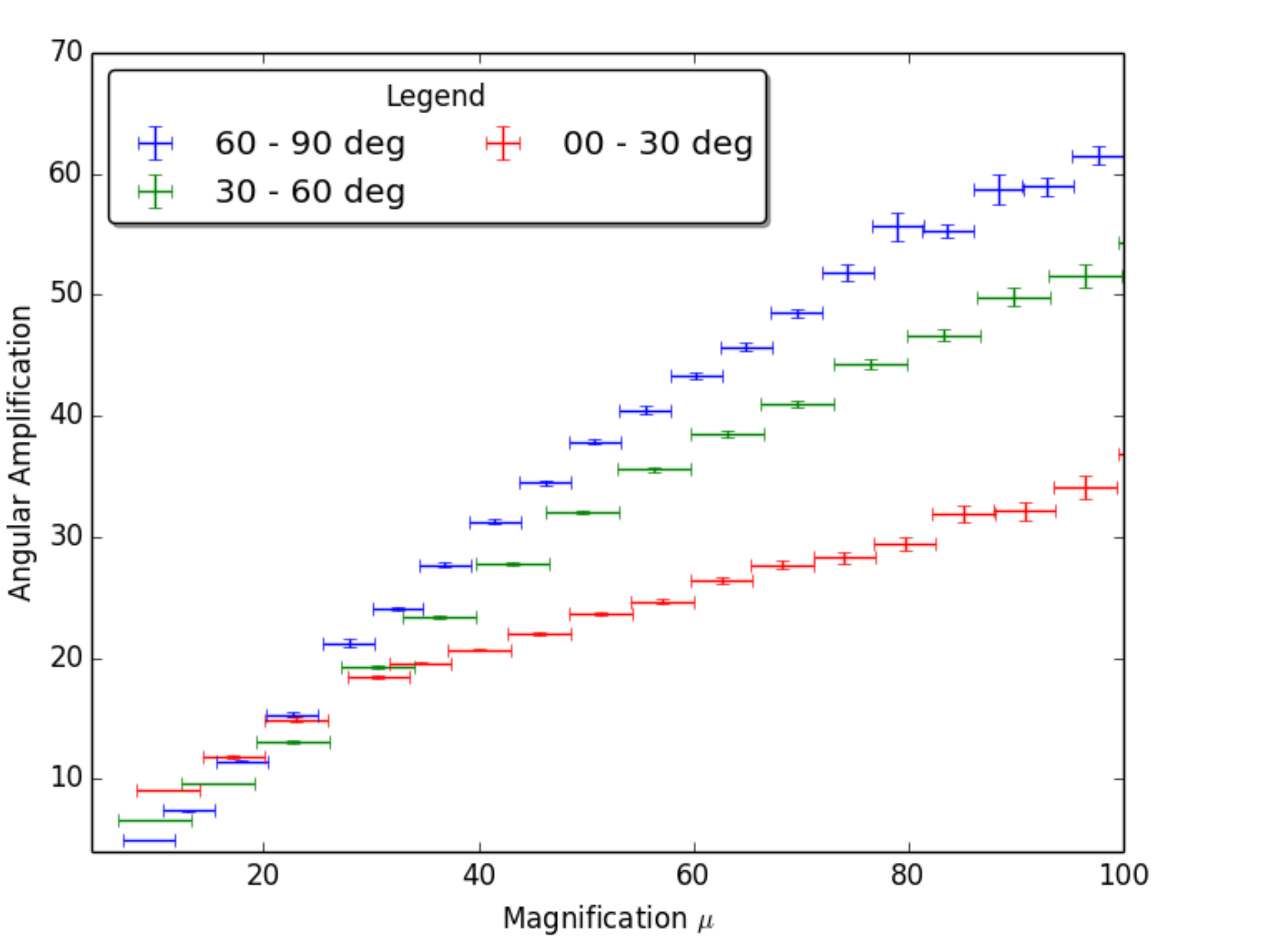}
\end{center}
\caption{\label{fig:AA}  
                          Offset amplification as a function of total magnification for different  angles of pair of sources in relation to the caustic. }
\end{figure}

For every selected source position, we pick a randomly oriented point at 1~or $5\,mas$ distance. 
This step allows us to investigate all possible orientations of the pairs of points in relation to the caustic. 
For such selected pairs of sources, we obtain the positions of mirage images. 
Next, we calculate the total offset between these positions of mirage images, 
and we normalize it by the distance between the pair of sources to obtain the offset amplification.
The offset amplification as  a function of total magnification is shown in Figure~\ref{fig:AA}.

The largest offset amplification  is achieved when the pair of sources is located perpendicularly  to the caustic. 
When the pair of sources is located within $1\% \,r_E$ from the caustic, 
the flux magnification is $\sim 70$, and the offset amplification reaches 50. 

\section{Probability of a Caustic Configuration}
\label{sec:CausticProbability} 
 
A caustic configuration, 
in which an angular offset in the source plane is amplified dozen times in the position of mirage images,
 requires the source to be located within a few percent of the Einstein radius inside the caustic. 
 The cross-section for a caustic region is small in comparison to the area within the Einstein radius. 
 
We evaluate the cross-section for the caustic region for an elliptical lens with parameters as described in Section~\ref{sec:ToyModel}.
The caustic length is $\sim1.15\,r_E$. 
Next, we require that the strongly lensed source must be located within an  $0.02\,r_E$ from the caustic to achieve significant offset amplification.
In such a configuration, the probability that the lensed source will be located in the caustic configuration is $\sim0.005$. 
 
\begin{figure}
\begin{center}
\includegraphics[width=11.cm,angle=0]{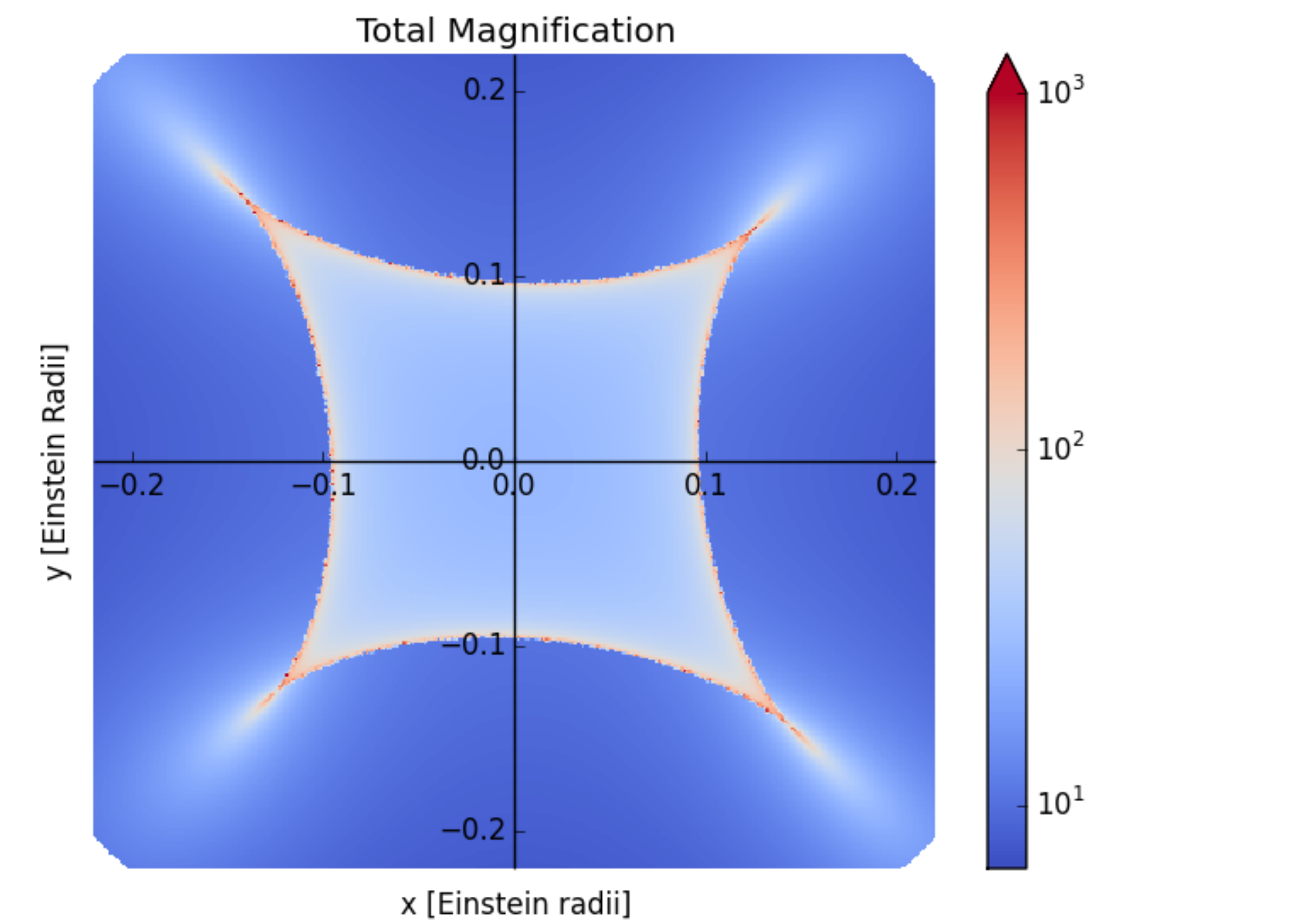}
\end{center}
\caption{\label{fig:TotalMag}  
                          Total flux magnification defined as a sum of flux magnifications of all mirage images of  a source. 
                          Coordinates are shown relative to the lens center. }
\end{figure}

 However, in the caustic configuration, sources are highly magnified. 
 The magnification of the source close to the caustic is greater than 20 (see Figure~\ref{fig:TotalMag}). 
 This introduces magnification bias that increases a probability of observing gravitationally lensed systems in the caustic configuration
 \citep{1980ApJ...242L.135T,1984ApJ...284....1T,2003ApJ...583...58W,2002ApJ...577...57W}. 
 We use Monte Carlo simulations to evaluate the magnification bias for the caustic region. 
 We calculated the magnification along the caustic region, 
 which we normalize to the integrated magnification within the Einstein radius. 
 We obtain 2\% probability that the source will be located within the region with magnification greater than 20. 
 However, due to the magnification bias, the probability of observing the lens system in the caustic configuration increases to 8\%. 
 Thus, a significant fraction of the observed gravitationally lensed quasars should be in the caustic configuration. 
 
 Due to the high magnification along the caustic, 
 the brightest observed lensed sources are more likely to be in the caustic configuration.  
 The magnification  bias can be seen in the Cosmic Lens All-Sky Survey (CLASS) and the Jodrell/VLA Astrometric Survey (JVAS) surveys that list 20 flat-spectrum radio sources. 
 Remarkably, about 8 sources out of 20 in the CLASS/JVAS surveys have mirage image morphology characteristic of sources in the caustic configuration. 
 
The magnification bias will also allow us to detect a population of very faint and distant quasars. 
The magnification close to the caustic as high as $\sim 100$ will allow us to detect and resolve the formation of  the first quasars. 
 
 \section{Astrometry}
\label{sec:Astrometry} 

The precise determination of the offsets between the radio and optical emission relies on accurate astrometry. 
For example, the HST absolute astrometry has a typical uncertainty of 0.2"-1", 
which is greater than the angular resolution. 
The HST fine guidance sensors operate by locking on individual stars,  
the resulting absolute astrometry associated with HST images is limited by
the accuracy of positions of individual Guide Star Catalog \citep{2008AJ....136..735L} stars.
The accuracy of absolute astrometry may be considerably improved by matching multiple objects visible on HST images
to deep ground-based astrometric catalogs \citep{2016AJ....151..134W}.
HST does provide excellent relative astrometry for sources located in crowded fields with a large field of view,
but this is not the case for observations of radio-loud quasars.
Inaccurate astrometry can introduce a systematic offset between optical and radio emission.

However, in the caustic configuration, four images of a source are observed. 
The multiple images provide a reference frame that can be used to find radio and optical offsets independent of the coordinate system. 
The positions of mirage images are measured in relation to the brightest  image,
the source position is reconstructed using a lens modeling and can be defined in relation to the positions of mirage images, or center of the lens.
The relative positions of point-like sources on HST images can be measured with sub-mas accuracy \citep{2011PASP..123..622B}.
Thus, the angular offset can be determined with even sub-mas accuracy  in relation to the caustic of the lensing galaxy 
and can be converted into physical units knowing redshift of the source.
The presence of a gravitational lens eliminates the need for absolute astrometry.  
As such, the caustic configuration will reduce possible systematics in the offset measurement arising from imprecise astrometry.

\section{Discussion}
\label{sec:Discussion} 

We evaluate the offset amplification for lensed sources located close to a caustic. 
In this section, we advocate using the gravitational offset amplification to elucidate the spatial origin of radio emission in relation to shorter wavelengths.

 \subsection{Physical Origin of Multi-wavelength Emission}
\label{sec:RadioVSOptical} 

Quasars powering extragalactic jets are complex objects with multiple sources and sites of emission \citep{1995PASP..107..803U,2000ApJ...545...63E}. 
Here, we discuss one of possible scenarios to explain offsets by considering the geometry of the innermost relativistic jets \citep{2016PASJ...68R...1H,2017ApJ...834...65A}.
The well-resolved radio observations of M87 revealed that 
 the jet is  maintaining the parabolic morphology from the base of the jet  up to
the HST-1 knot where it transitions to conical shape  \citep{2012ApJ...745L..28A,2013ApJ...775..118N}.

The radiation emitted by jets is relativistically Doppler boosted toward the observer by $\mathcal{D}^{n}$ \citep{1966Natur.211..468R,2003A&A...406..855M,2007ApJ...658..232C}. 
The Doppler factor is defined as $\mathcal{D}=[\Gamma(1-\beta \cos{\theta_{obs}})]^{-1}$,  
where $\Gamma=(1-\beta^2)^{-1/2}$ is the Lorenz factor,
$\beta=v/c$ is  the velocity of moving plasma, $v$, in units of the speed of light $c$,  
and  $\theta_{obs}$ is the angle to the line-of-sight with the observer. 
The exponent $n$ combines effects due to the K correction \citep{2002astro.ph.10394H} and the Doppler boosting cased by relativistic aberration, 
time dilation, and the solid angle transformation \citep{1995PASP..107..803U}.
In the following calculations, we assume $n=4$. 

The parabolic part of the M87 jet close to the central engine (within 1 pc) is not well collimated  \citep{1999Natur.401..891J,2013ApJ...775...70H}.
The apparent  jet opening angle of  M87 at  the distance of $0.1\,$pc is $\theta_{jet}=33^\circ$. 
Jets became well collimated at larger distances.
At the distance of the HST-1 knot, the jet opening angle is $ \theta_{jet}\sim6^\circ$. 
The Lorenz factor can be approximated as $1/\theta_{jet}$ \citep{2014ApJ...794L...8B}. 
The viewing angle of M87 is no more than $\theta_{obs}\leq19^\circ$ from our line-of-sight \citep{1999ApJ...520..621B}.
In such configuration, we expect Doppler factor of $\mathcal{D}\sim3$, for both, the emission close to the central engine and the HST-1 knot. 

If we imagine the M87 jet pointed toward us at the viewing angle of $\theta_{obs}\sim3^\circ$, as in the case for blazars,
the radiation originating from the region close to the supermassive black hole would have the Doppler factor of $\mathcal{D}\sim3$.
However, the HST-1 knot, due to its collimation, would have the Doppler factor of $\mathcal{D}\sim16$. 
Taking into calculations that the radiation is enhanced by $\mathcal{D}^4$, 
and assuming similar intrinsic luminosities, the HST-1 knot would appear $\sim500$ times brighter that the emission close to the central engine.

The HST-1 knot is located at a projected distance of $\sim60\,$pc from the supermassive black hole \citep{1999ApJ...520..621B}.
If M87 were located at redshift equals 1, the emission from the HST-1 knot would appear at an offset of $\sim7\,$mas from the supermassive black hole. 
Thus, relativistically boosted recollimation shocks are good candidates to explain offsets between radio and optical emission. 
The gravitationally lensed radio-loud quasars in the caustic configuration will allow us to investigate this scenario. 

The other possibility is that black holes powering AGN do not always reside at the centres of their host galaxies  \citep{2012MmSAI..83..925B}.
Such systems could host a binary black hole, or they could  be offset from the galaxy center after receiving  a kick from binary coalescence. 
Close binary systems cannot be directly resolved using optical telescopes.
However, such systems could be easily identified in the caustic configuration.
In the caustic configuration, the binary system would produce 8 mirage images if both black holes are located inside the caustic, 
or 6 images if one black hole is located inside the caustic and the second one is outside the caustic.

\subsection{Multi-wavelength Offset Observations}
\label{sec:MOO} 

The first comparison of Gaia and VLBI positions revealed that $\sim6\%$ of sources show significant offsets \citep{2016A&A...595A...5M}. 
The combination of offsets measured using Gaia and gravitational lensing will give us a complementary insight into inner regions of galaxies at all redshifts. 
It is also expected that Gaia can detect more than 500 000 quasars  \citep{2015A&A...574A..46P},
and among them  about 3000 gravitationally lensed quasars  \citep{2012MmSAI..83..944F}. 
Gaia astrometry for these gravitationally lensed quasars will provide an excellent frame for comparison of mirage image positions. 

The lensing probability depends on distance to a source. 
Thus, gravitational lensing is a powerful tool to investigate sources at high redshift and all frequencies from radio up to very-high-energy gamma rays.
Gaia provides a unique way to investigate optical emission of nearby AGN.  

The flux magnification in the caustic configuration will allow us to detect a population of faint quasars
and will enhance our capability to search for the most distant quasars. 
The identification of high redshift quasars with offsets between optical and radio emission will allow us to select sources 
for follow-up observations with the James Webb Space Telescope\footnote{https://jwst.nasa.gov} (JWST),  
the Wide Field Infrared Survey Telescope\footnote{https://wfirst.gsfc.nasa.gov} (WFIRST),
or ground facilities like the Extremely Large Telescope\footnote{http://www.eso.org/public/teles-instr/elt/} (ELT) equipped with adaptive optics. 

Moreover, the caustic configuration will allow us to resolve X-ray emission. 
The angular resolution of the Chandra satellite\footnote{http://chandra.harvard.edu} is $0.5"$,
which corresponds to $4\,$kpc at redshift $\sim1$. 
The M87 jet consists of bright knots of radio, optical, and X-ray emission spread throughout a projected distance of $\sim1.6\,$kpc. 
Thus, if M87 were located at redshift  $\sim1$ then the Chandra satellite would observe it as a point source. 
However, if M87-like source was gravitationally lensed in the caustic configuration, 
the offset amplification of 50 in combination with advantage of relative astrometry 
would allow us not only to resolve the jet, but also separate HST-1-like structures from the supermassive black hole. 

The future X-ray missions including Lynx\footnote{https://wwwastro.msfc.nasa.gov/lynx/} and {\tt ATHENA}\footnote{http://www.the-athena-x-ray-observatory.eu}
will not provide an improvement in angular resolution. 
Thus, the gravitational lensing is the only way to resolve the origin of the X-ray emission at scales smaller than  $0.5"$.

\subsection{Lens Modeling}
\label{sec:Modeling} 

We find that a significant fraction of gravitationally lensed quasars will be in the caustic configuration. 
Many of these sources already have  archival radio and optical observations,
and have well-reconstructed models of their lenses.
Moreover, the lensing potential near a critical curve where images form when the source is close to the caustic
can be extracted based on the generic properties of images using the model independent approach proposed by \citet{2016A&A...590A..34W} and \citet{2016arXiv161201793W}.
Thus, the analysis of these sources will give a foundation for using gravitationally lensed quasars in the caustic configuration 
to elucidate the origin of radio emission in respect to optical emission.
This study will shed new light on our understanding of processes responsible for particle acceleration at large
distances from supermassive black holes.

\subsection{Euclid and SKA Synergy}
\label{sec:Future} 

Our method relies on having accurate positions of mirage images at multiple wavelengths. 
Thus, the Square Kilometre Array  (SKA) and Euclid will provide the perfect set of observations to apply the method to a large ensemble of sources.  
SKA will provide observations with a resolution of $\sim2\,mas$ at $10\,$GHz, and $\sim20\,mas$
at $1\,$GHz \citep{2009IEEEP..97.1482D,2012PASA...29...42G}. 
The well-resolved radio positions of mirage images of gravitationally lensed quasars 
will set a foundation for reconstructing the mass distribution of lenses, 
and it will provide a reference frame for comparison with other observations.

Euclid will map three-quarters of the extragalactic sky with Hubble Space Telescope resolution to $\sim24\,$mag \citep{2011arXiv1110.3193L,2013LRR....16....6A}. 
It is expected that  Euclid and SKA might detect $\sim10^5$ gravitationally lensed compact flat-spectrum AGN \citep{2004NewAR..48.1085K,2015aska.confE..84M,2016arXiv160400271S}.
A significant fraction of these sources will be in the caustic configuration.
These observations will allow us to investigate the offset distribution for a  large ensemble of sources 
and so a statistical investigation of the origin of emission from extragalactic jets.

\section{Conclusions}
\label{sec:Conclusions}

Our ability to study the multi-wavelength structures of distant quasars is limited by the insufficient angular resolution of current telescopes and astrometry.
We demonstrate using Monte Carlo simulations that when a pair of sources with an angular offset is located close to the caustic of a lensing galaxy,
then the offset in the position of the sources  is amplified by a factor of tens in the image plane. 
Thus, if the multi-wavelength emissions originate from the same region, 
the position of mirage images will be consistent across the entire electromagnetic spectrum. 
However, if the source structure is complex with spatial offsets between radio and optical emission of even $\sim1\,mas$ ($\sim10\,$pc),
then the positions of mirage images can differ by even more than $\sim50\,mas$.
The difference in the position of the mirage images can be used to evaluate offsets between emitting regions.

We find that when the source is located within 1\% Einstein radius from the caustic, 
the offset amplification is greater than 50. 
Thus, the gravitationally lensed sources in the caustic configuration can be used to elucidate the origin of multi-wavelength emission on milliarcsecond scales using existing facilities. 

The magnification bias significantly boosts the probability of observing gravitationally lensed quasars in the caustic configuration. 
We find that due to the magnification bias, even 8\% of observed gravitationally lensed quasars could be in the caustic configuration.

The ability to zoom into milliarcsecond scales with multi-wavelength observations combined with flux magnification
will allow us to investigate phenomena related to
supermassive black holes that include relativistic jets,  recoiling supermassive black holes, or binary systems.
In the foreseeable future, the synergy between SKA and Euclid surveys will provide observations of hundreds of thousands of strongly lensed sources. 
These observations will allow us to elucidate the spatial origins of radio and optical emissions for a large ensemble of sources.
The offset distribution will enable us to explain the physical origins of radio emission, 
their relation to supermassive black holes, as well as their cosmic evolution including the formation of the first quasars.

\acknowledgments
I thank the referee for providing very valuable comments which greatly helped to improve the manuscript. 
I would like to thank Martin Elvis, Margaret Geller, Jan Kansky, Robert Kirshner, Mike McCourt, Masanori Nakamura, Michal Ostrowski, 
Bronek Rudak, Aneta Siemiginowska, and Dan Schwartz for comments and useful discussions.

A.B. is  supported by NASA through an Einstein Postdoctoral Fellowship.
This research was supported in part by PLGrid Infrastructure.

\appendix

\section{Gravitational Lensing: Formalism}
\label{sec:Formalism}

We begin with a brief introduction to  gravitational lensing. 
Positions of the source, $\vec{\beta}$,  and the mirage images, $\vec{\theta}$, are related through the lens equation 
\begin{equation}
\vec{\beta} = \vec{\theta} -\vec{\alpha}(\vec{\beta} ) \,,
\end{equation}
where $\alpha$ is the deflection angle. 
The lens equation is not linear. 
Thus, creation of multiple mirage images is expected for a single source position. 

The deflection angle $\alpha$ is the gradient of the effective gravitational potential $\psi$
\begin{equation}
\vec{\alpha} = \vec{\triangledown}_\theta\psi \,,
\end{equation}
while the Laplacian of the gravitational potential is proportional to the surface-mass density $\Sigma$
\begin{equation}
\triangledown^2_\theta\psi = \frac{2}{c^2}  \frac{D_{LS}D_{OL}}{D_{OS}}4\pi G \Sigma = 2 \frac{\Sigma(\vec{\theta})}{\Sigma_{cr}} \equiv 2 \kappa (\vec{\theta})\,,
\end{equation}
where $D_{LS},D_{OL}$, and $D_{OS}$ are the angular diameter distances
from the lens to the source,
from the observer to the lens, 
 and from the observer to the source, respectively.
The convergence $\kappa(\vec{\theta})$ is the surface mass density scaled with its critical value $\Sigma_{cr}$ defined as
\begin{equation}
\Sigma_{cr} = \frac{c^2}{4\pi G} \frac{D_{OS}}{D_{LS}D_{OL}} \,.
\end{equation}

The properties of the lens mapping from the source to the lens plane are described by the Jacobian matrix A
\begin{equation}
A \equiv \frac{\partial \vec{\beta}}{\partial \vec{\theta}} = \left( \delta_{ij} - \frac{\partial \alpha_i(\vec{\theta}) }{\partial \theta_j} \right) \,.
\end{equation} 

The Jacobian A is in general a function of position $\vec{\theta}$, 
and is used to calculate the magnification, $\mu$, as an inverse of the determinant of A
\begin{equation} 
\mu = \frac{1}{\mbox{det}A} \,. 
\end{equation}

In geometrical optics approximation, the determinant of A, $\mbox{det}A=0$, corresponds to infinite magnification. 
In the lens plane, points where the magnification goes to infinity are called critical curves. 
These critical curves define regions where mirage images merge or are created. 

The critical curves mapped to the source plane are called caustics. 
When the source crosses a caustic curve, the number of mirage images changes. 
In the case of a single point lens, the caustic degenerates into a point. 
For a spherically symmetric mass distribution, the critical curves are circles. 
For elliptical lenses or spherically symmetric lenses plus external shear, the caustics can consist of cusps and folds 
\citep{1997MNRAS.292..863W,1986ApJ...310..568B,2006MNRAS.372.1692A}.
The shape of caustic can be distorted by substructure in the lensing galaxy \citep{2006glsw.conf.....M}.

If the source is located relatively close to the caustic,
then even small changes of the position of the source can produce large changes in magnifications and positions of the mirage images.
Mathematically, the approximate of the caustic is best described by catastrophe theory and Morse theory \citep{1993A&A...268..453E}.

\begin{figure}
\begin{center}
\includegraphics[width=8.5cm,angle=0]{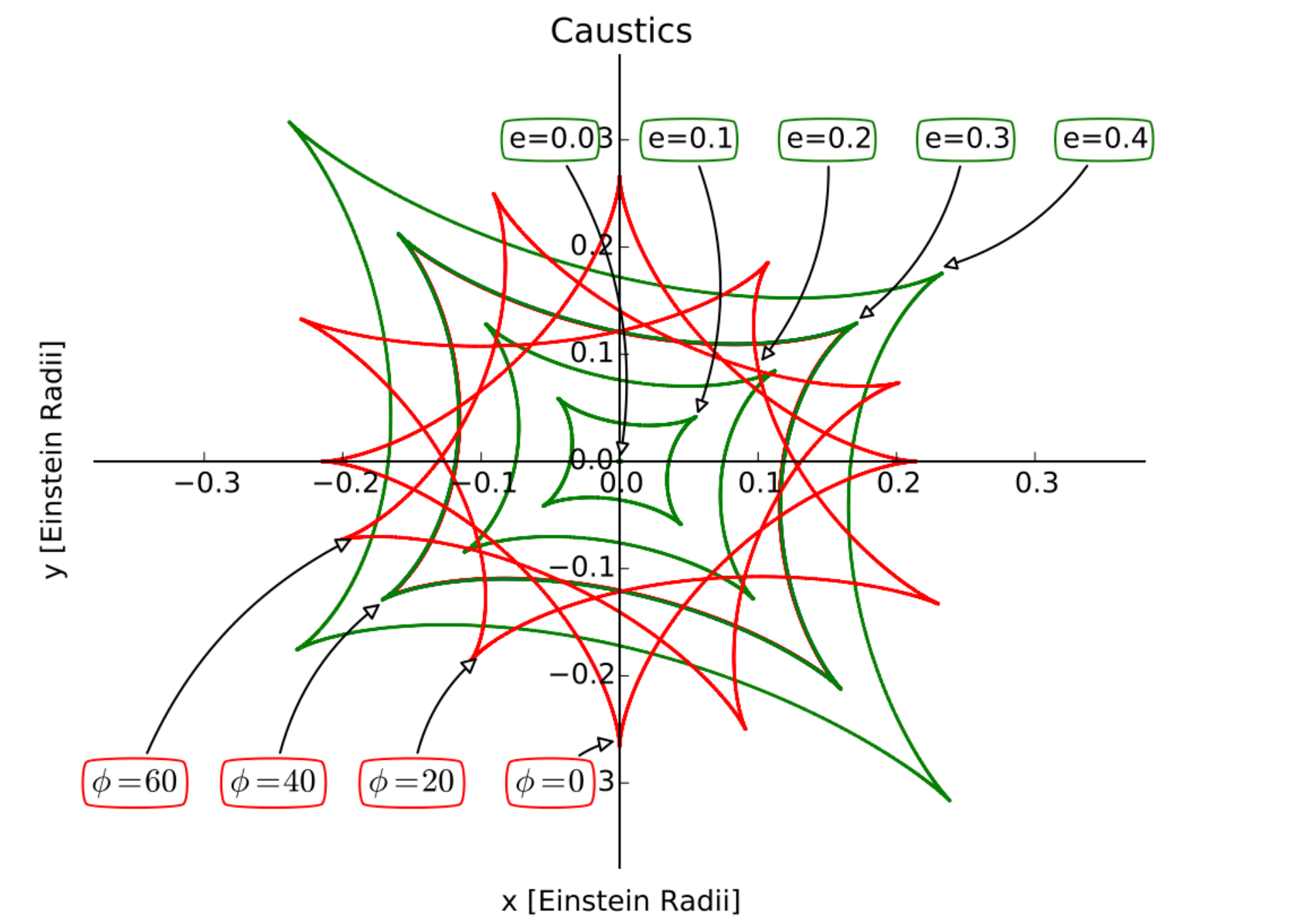}
\end{center}
\caption{\label{fig:Caustic} 
                          Caustics for a singular isothermal ellipsoid mass distribution in units of the Einstein ring radius.  
                          Green lines represent caustics for a range of ellipticities and a fixed angle $\phi=40^\circ$.
                          Red lines indicate caustics for a range of angles, and a fixed ellipticity of $e=0.3$. 
                          }
\end{figure}

\section{Caustic of an Elliptical Lens}
\label{sec:Caustic}

Different reflective or refractive surfaces can produce a variety of caustic curves.
Here, we investigate caustics created by an elliptical lens.
The elliptical lenses provide best approximation of the gravitational potential of a typical galaxy. 

The model that is the most frequently used to describe a lensing galaxy is the Singular Isothermal Ellipsoid (SIE).
The SIE has the three-dimensional radial profile of $\rho \propto r^{-2}$, and a convergence given by
\begin{equation}
\kappa = \frac{b_{SIE}(q)}{2\sqrt{\tilde{x}^2 +\tilde{y}^2/q^2}} \,,
\end{equation}
where $e$ is an ellipticity, and the axis ratio is $q=1-e$, with $q=1$  for a spherical case. 
The normalization factor $b_{SIE}$ is related to the velocity dispersion, $\sigma$, as:
\begin{equation}
b_{SIE}(1) = 4\pi \left( \frac{\sigma}{c} \right)^2 \frac{D_{LS}}{D_{OS}} \,,
\end{equation}
where 
 the coordinates $\tilde{x}$ and $\tilde{y}$ are rotated by an angle $\phi$
 \begin{equation}
\begin{split}
 \tilde{x}  = x \cos \phi + y \sin \phi \,,  \\
 \tilde{y} =  -x \sin \phi + y \cos \phi \,.
\end{split}
 \end{equation}

Figure~\ref{fig:Caustic} shows caustics resulting from deflection of rays in the spacetime curved by an elliptical galaxy located at redshift equals 1.
In this example, the source is located at redshift equal 2. 
Caustics are obtained using {\tt glafic} code. 
The caustic size scales with an Einstein radius of the lens. 
The angle $\phi$ does change the orientation of the caustic, 
but has no influence on the caustic shape or size (see Figure~\ref{fig:Caustic}). 

For SIE with an ellipticity $e=0$, 
the caustic is represented by a point at the center of the lens. 
The lens ellipticity defines the caustic length. 
Lenses with greater ellipticity have larger caustics. 
Thus, the probability that a background source will be located close to the caustic increases with the lens ellipticity
and the Einstein ring radius.

\bibliography{TurningGalaxies}
\end{document}